# Integrating R and Hadoop for Big Data Analysis

**Bogdan OANCEA** (bogdanoancea@univnt.ro)
"Nicolae Titulescu" University of Bucharest

**Raluca Mariana DRAGOESCU** (dragoescuraluca@gmail.com)
The Bucharest University of Economic Studies

### ABSTRACT

*Analyzing and working with big data could be very difficult using classical means like relational database management systems or desktop software packages for statistics and visualization. Instead, big data requires large clusters with hundreds or even thousands of computing nodes. Official statistics is increasingly considering big data for deriving new statistics because big data sources could produce more relevant and timely statistics than traditional sources. One of the software tools successfully and wide spread used for storage and processing of big data sets on clusters of commodity hardware is Hadoop. Hadoop framework contains libraries, a distributed file-system (HDFS), a resource-management platform and implements a version of the MapReduce programming model for large scale data processing. In this paper we investigate the possibilities of integrating Hadoop with R which is a popular software used for statistical computing and data visualization. We present three ways of integrating them: R with Streaming, Rhipe and RHadoop and we emphasize the advantages and disadvantages of each solution.*
  **Keywords:** *R, big data, Hadoop, Rhipe, RHadoop, Streaming*
  ***JEL Classification:** L8, C88*

## 1. INTRODUCTION

  The big data revolution will transform the way we understand the surrounding economic or social processes. We can no longer ignore the enormous volume of data being produced every day. The term "big data" was defined as data sets of increasing volume, velocity and variety (Mayer-Schönberger, 2012), (Beyer, 2011). Big data sizes are ranging from a few hundreds terabytes to many petabytes of data in a single data set. Such amount of data is hard to be managed and processed with classical relational database management systems and statistics and visualization software packages – it requires high computing power and large storage devices.
  Official statistics need to harness the potential of big data to derive more relevant and timely statistics but this not an easy process. The first step is to identify the sources of big data possible to be used in official statistics. According to (HLG, 2013) large data sources that can be used in official statistics are:
- Administrative data;
- Commercial or transactional data, such as on-line transactions using credit cards;
- Data provided by sensors (satellite imaging, climate sensors, etc.);
- Data provided by tracking devices (GPS, mobile devices, etc.);
- Behavioral data (for example Internet searches);
- Data provided by social media.

Using big data in official statistics raises several challenges (HLG, 2013). Among them we ca mention: legislative issues, maintaining the privacy of the data, financial problems regarding the cost of sourcing data, data quality and suitability of statistical methods and technological challenges. At this time there are several international initiatives that try to outline an action plan for using Big Data in official statistics: Eurostat Task Force on Big Data, UNECE's Big data HLG project.

At this time there are some ongoing projects that already used big data for developing new statistics implemented by statistical agencies. We can mention (HLG, 2013):
- Traffic and transport statistics computed by Statistics Netherlands using traffic loop detection records generated every day. There are 10,000 detection loops on Dutch roads that produce 100 million records every day;
- Social media statistics computed also by Statistics Netherlands. Dutch Twitter produces around 1 million public social media messages on a daily basis. These messages were analyzed from the perspective of content and sentiment;
- The software developed at Eurostat for price scrapping from the Internet to assist in computing the Consumer Price Index;
- The Billion project developed at MIT (http://bpp.mit.edu/) is a project that collect prices from retailers around the world to conduct economic research;
- Tourism Statistics developed in Estonia by using mobile positioning data (Ahas, 2013);

In this paper we will investigate a technological problem – we will present a way of integrating Hadoop (White, 2012), a software framework for distributed computing used for big data processing with R (R Core Team, 2013) which is a popular statistics desktop package. The paper is structured as follows: in section 2 we will introduce R and Hadoop. Next we will show how these two software packages can be integrated in order to be used together for big data statistical analysis and section 4 will conclude our presentation.

## 2. R AND HADOOP - SOFTWARE TOOLS FOR LARGE DATA SETS STATISTICAL ANALYSIS

R is a free software package for statistics and data visualization. It is available for UNIX, Windows and MacOS platforms and is the result of the work of many programmers from around the world. R contains facilities for data handling, provides high performance procedures for matrix computations, a large collection of tools for data analysis, graphical functions for data visualization and a straightforward programming language. R comes with about 25 standard packages and many more packages available for download through the CRAN family of Internet sites (http://CRAN.R-project.org). R is used as a computational platform for regular statistics production in many official statistics agencies (Todorov, 2010), (Todorov, 2012). Besides official statistics, it is used in many other sectors like finance, retail, manufacturing, academic research etc., making it a popular tool among statisticians and researchers.

Hadoop is a free software framework developed with the purpose of distributed processing of large data sets using clusters of commodity hardware, implementing simple programming models (White, 2013). It is a middleware platform that manages a cluster of computers that was developed in Java and although Java is main programming language for Hadoop other languages could be used to: R, Python or Ruby. Hadoop is available at http://hadoop.apache.org/. One of the biggest users of Hadoop is Yahoo!. Yahoo! uses Hadoop for the *Yahoo! Search Webmap* which is an application that runs on a very large cluster and produces data used in Yahoo! Web search queries (Yahoo! Developer Network, 2014). Another Hadoop important user is Facebook that operated a Hadoop cluster with more than 100 PB of data in 2012 (Ryan, 2012).

The Hadoop framework includes:

- Hadoop Distributed File System (HDFS) - a high performance distributed file system;
- Hadoop YARN which is a framework for job scheduling and cluster resource management;
- Hadoop MapReduce – a system for parallel processing of large data sets that implements the MapReduce model of distributed programming (Dean, 2004).

In brief, Hadoop provides a reliable distributed storage through HDFS and an analysis system by MapReduce. It was designed to scale up from a few servers to hundreds or thousands of computers, having a high degree of fault tolerance. Hadoop is now a de-facto standard in big data processing and storage, it provides unlimited scalability and is supported by major vendors in the software industry.

Hadoop Distributed File System relies on a client/server architecture consisting in a single NameNode implemented on a master server that manages the file system namespace and a number of DataNodes which manage storage attached to the nodes. Files are split into blocks that are stored in a set of DataNodes. The NameNode is responsible with operations like opening, closing or renaming files while DataNodes are responsible for responding to read or write requests from clients.

MapReduce is a model for processing large sets of data in-parallel on large clusters computers. It splits the input data in chucks that are processed in parallel by the *map tasks*. The results of the map tasks are sorted and forwarded as inputs to the *reduce tasks* that performs a summary operation. The framework that implements the MapReduce paradigm should marshal the distributed servers, run tasks in parallel, manage the data transfers between the nodes of the cluster, and provide fault tolerance. Hadoop MapReduce hides the parallelism from the programmer, presenting him a simple model of computation.

The main features of the Hadoop framework can be summarized as follows:
- High degree of scalability: new nodes can be added to a Hadoop cluster as needed without changing data formats, or application that runs on top of the FS;
- Cost effective: it allows for massively parallel computing using commodity hardware;
- Flexibility: Hadoop differs from RDBMS, being able to use any type of data, structured or not;
- Fault tolerance: if a node fails from different reasons, the system sends the job to another location of the data and continues processing.

Hadoop has also a series of limitations which can be summarized as follows:

- HDFS is an append-only file system, it doesn't allow update operations;

- MapReduce jobs run in batch mode. That's why Hadoop is not suited for interactive applications;
- Hadoop cannot be used in transactional applications.

Data analysts who work with Hadoop may have a lot of R scripts/packages that they use for data processing. Using these scripts/packages with Hadoop normally requires rewriting them in Java or other language that implements MapReduce. This is cumbersome and could be a difficult and error prone task. What we need is a way to connect Hadoop with R and use the software already written for R with the data stored in Hadoop (Holmes, 2012). Another reason for integrating R with Hadoop for large data sets analysis is the way R works – it processes the data loaded in the main memory. Very large data sets (TB or PB) cannot be loaded in the RAM memory and for these data Hadoop integrated with R is one of the first choice solutions. Although there are many solutions for using R on a high performance computing environment ( snow, rmpi or rsge) all these solutions require that the data must be loaded in memory before the distribution to computing nodes and this is simple not possible for very large data sets.

## 3. R AND HADOOP INTEGRATION

We will present three approaches to integrate R and Hadoop: R and Streaming, Rhipe and RHadoop. There are also other approaches to integrate R and Hadoop. For example RODBC/RJDBC could be used to access data from R but a survey on Internet shows that the most used approaches for linking R and Hadoop are Streaming, Rhipe (Cleveland, 2010) and RHadoop (Prajapati, 2013).

The general structure of the analytics tools integrated with Hadoop can be viewed as a layered architecture presented in figure 1.

The first layer is the hardware layer – it consists in a cluster of (commodity) computers. The second layer is the middleware layer – Hadoop. It manages the distributions of the files by using HDFS and the MapReduce jobs. Then it comes a layer that provides an interface for data analysis. At this level we can have a tool like Pig which is a high-level platform for creating MapReduce programs using a language called Pig-Latin. We can also have Hive which is a data warehouse infrastructure developed by Apache and built on top of Hadoop. Hive provides facilities for running queries and data analysis using an SQL-like language called HiveQL and it also provides support for implementing MapReduce tasks.

Besides these two tools we can implement at this level an interface with other statistical software like R. We can use Rhipe or Rhadoop libraries that build an interface between Hadoop and R, allowing users to access data from the Hadoop file system and write their own scripts for implementing Map and Reduce jobs, or we can use Streaming that is a technology integrated in Hadoop.

Hadoop can be also integrated with other statistical software like SAS or SPSS.

**Hadoop and data analysis tools**

*Figure 1*

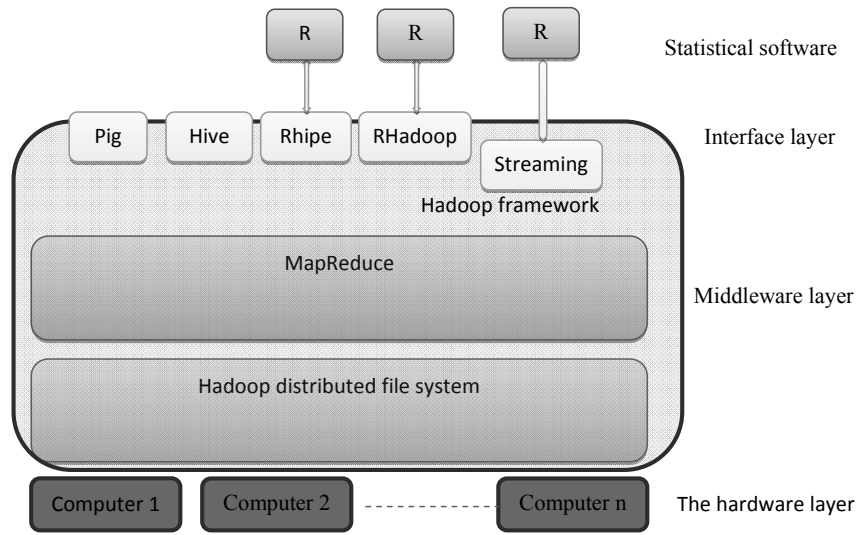

We will analyze these options of integration between R and Hadoop from different point of views: licensing, complexity of installation, benefits and limitations.

## R AND STREAMING

Streaming is a technology integrated in the Hadoop distribution that allows users to run Map/Reduce jobs with any script or executable that reads data from standard input and writes the results to standard output as the mapper or reducer. This means that we can use Streaming together with R scripts in the map and/or reduce phase since R can read/write data from/to standard input. In this approach there is no client-side integration with R because the user will use the Hadoop command line to launch the Streaming jobs with the arguments specifying the mapper and reducer R scripts.

A command line with map and reduce tasks implemented as R scripts would look like this:

**An example of a map-reduce task with R and Hadoop integrated by Streaming framework**

*Figure 2*

```
$ ${HADOOP_HOME}/bin/Hadoop jar
${HADOOP_HOME}/contrib/streaming/*.jar \
-inputformat org.apache.hadoop.mapred.TextInputFormat \
-input input_data.txt \
-output output \
-mapper /home/tst/src/map.R \
-reducer /home/tst/src/reduce.R \
-file /home/tst/src/map.R \
-file /home/tst/src/reduce.R
```

Here we supposed that the data in the file "input_data.txt" was already copied from the local file system to the HDFS. The meaning of the command line parameters are:

"-inputformat org.apache.hadoop.mapred.TextInputFormat" specifies the input format for the job (we stored our input data in a text file);

"-input input_data.txt" specifies the input data file of our job;

"-output output" sets the output directory of the job;

"-mapper /home/tst/src/map.R" specifies the map phase executable. In this example we used an R script named map.R located in /home/tst/src/ directory;

"-reducer /home/tst/src/reduce.R" specifies the reduce phase executable. In our example, the reducer is also an R script named reduce.R located in /home/tst/src/ directory;

"-file /home/tst/src/map.R" indicates that the R script map.R should be copied to the distributed cache, made available to the map tasks and causes the map.R script to be transferred to the cluster machines where the map-reduce job will be run;

"-file /home/tst/src/reduce.R" indicates that the R script reduce.R should be copied to the distributed cache an made available to the map tasks and causes the reduce.R script to be transffered to the cluster machines where the map-reduce job will be run.

The integration of R and Hadoop using Streaming is an easy task because the user only needs to run Hadoop command line to launch the Streaming job specifying the mapper and reducer scripts as command line arguments. This approach requires that R should be installed on every DataNode of the Hadoop cluster but this is simple task.

The licensing scheme need for this approach implies an Apache 2.0 license for Hadoop and a combination of GPL-2 and GPL-3 for R.

## RHIPE

Rhipe stands for "R and Hadoop Integrated Programming Environment" and is an open source project that provides a tight integration between R and Hadoop. It allows the user to carry out data analysis of big data directly in R, providing R users the same facilities of Hadoop as Java developers have. The software package is freely available for download at www.datadr.org.

The installation of the Rhipe is somehow a difficult task. On each DataNode the user should install R, Protocol Buffers and Rhipe and this is not an easy task: it requires that R should be built as a shared library on each node, the Google Protocol Buffers to be built and installed on each node and to install the Rhipe itself. The Protocol Buffers are needed for data serialization, increasing the efficiency and providing interoperability with other languages.

The Rhipe is an R library which allows running a MapReduce job within R. The user should write specific native R `map` and `reduce` functions and Rhipe will manage the rest: it will transfer them and invoke them from map and reduce tasks. The map and reduce inputs are transferred using a Protocol Buffer encoding scheme to a Rhipe C library which uses R to call the map and reduce functions. The advantages of using Rhipe and not the parallel R packages consist in its integration with Hadoop that provides a data distribution scheme using Hadoop distributed file system across a cluster of computers that tries to optimize the processor usage and provides fault tolerance.

The general structure of an R script that uses Rhipe is shown in figure 3 and one can easily note that writing such a script is very simple.

**The structure of an R script using Rhipe**

*Figure 3*

```
1 library(Rhipe)
2 rhinit(TRUE, TRUE);

3   map<-expression   (   {lapply   (map.values,
function(mapper)…)})

4 reduce<-expression(
5 pre = {…},
6 reduce = {…},
7 post = {…},
8 )

9 x <- rhmr(
10           map=map, reduce=reduce,
11           ifolder=inputPath,
12           ofolder=outputPath,
13           inout=c('text', 'text'),
14           jobname='a job name'))

15 rhex(z)
```

      The script should begin with loading the Rhipe library into memory (line 1) and initializing the Rhipe library (line 2). Line 3 defines the map expression to be executed by the Map task. Lines 4 to 8 defines the reduce expression. Lines 4 to 8 define the reduce expression consisting in three callbacks. The `pre` block of instructions (line 5) is called for each unique map output key before these values being sent to the reduce block. The `reduce` block (line 6) is called then with a vector of values as argument and in the end, the `post` block (line 7) is called to emit the output (key, value) pair. Line 9 shows the call of `rhmr` function which set up the job (creates a MapReduce object) and `rhex` function call (line 15) that launches the MapReduce job to the Hadoop framework. Rhipe also provides functions to communicate with Hadoop during the MapReduce process like `rhcollect` that allows writing data to Hadoop MapReduce or `rhstatus` that returns the status of the a job.

      Rhipe let the user to focus on data processing algorithms and the difficulties of distributing data and computations across a cluster of computers are handled by the Rhipe and library and Hadoop.

The licensing scheme needed for this approach implies an Apache 2.0 license for Hadoop and Rhipe and a combination of GPL-2 and GPL-3 for R.

## RHADOOP

RHadoop is an open source project developed by Revolution Analytics (http://www.revolutionanalytics.com/) that provides client-side integration of R and Hadoop. It allows running a MapReduce jobs within R just like Rhipe and consist in a collection of four R packages:
- `plyrmr` - plyr-like data processing for structured data, providing common data manipulation operations on very large data sets managed by Hadoop;
- `rmr` – a collection of functions providing and integration of R and MapReduce model of computation;
- `rdfs` – an interface between R and HDFS, providing file management operations within R;
- `rhbase` - an interface between R and HBase providing database management functions for HBase within R;

Setting up RHadoop is not a complicated task although RHadoop has dependencies on other R packages. Working with RHadoop implies to install R and RHadoop packages with dependencies on each Data node of the Hadoop cluster. RHadoop has a wrapper R script called from Streaming that calls user defined map and reduce R functions. RHadoop works similarly to Rhipe allowing user to define the `map` and `reduce` operation. A script that uses RHadoop looks like:

**The structure of an R script using RHadoop**

*Figure 4*

```
1     library(rmr)
2     map<-function(k,v) { …}
3     reduce<-function(k,vv) { …}
4     mapreduce(
input ="data.txt",
               output="output",
               textinputformat =rawtextinputformat,
map = map,
reduce=reduce
)
```

First, the `rmr` library is loaded into memory (line 1) and then follows the definition of the `map` function which receives a (key,value) pair as input. The `reduce` function (line 3) is called with a key and a list of values as arguments for each unique map key. Finally, the scripts sets up and run the `mapreduce` job (line 4). It should be noted that `rmr` makes the client-side R environment available for `map` and `reduce` functions. The licensing scheme needed for this approach implies an Apache 2.0 license for Hadoop and RHadoop and a combination of GPL-2 and GPL-3 for R.

## 4. CONCLUSIONS

Official statistics is increasingly considering big data for building new statistics because its potential to produce more relevant and timely statistics than traditional data sources. One of the software tools successfully used for storage and processing of big data sets on clusters of commodity hardware is Hadoop. In this paper we presented three ways of integrating R and Hadoop for processing large scale data sets: R and Streaming, Rhipe and RHadoop. We have to mention that there are also other ways of integrating them like ROBDC, RJBDC or Rhive but they have some limitations. Each of the approaches presented here has benefits and limitations. While using R with Streaming raises no problems regarding installation, Rhipe and RHadoop requires some effort in order to set up the cluster. The integration with R from the client side part is high for Rhipe and Rhadoop and is missing for R and Streaming. Rhipe and RHadoop allows users to define and call their own `map` and `reduce` functions within R while Streaming uses a command line approach where the `map` and `reduce` functions are passed as arguments. Regarding the licensing scheme, all three approaches require GPL-2 and GPL-3 for R and Apache 2.0 for Hadoop, Streaming, Rhipe and RHadoop.

We have to mention that there are other alternatives for large scale data analysis: Apache Mahout, Apache Hive, commercial versions of R provided by Revolution Analytics, Segue framework or ORCH, an Oracle connector for R but Hadoop with R seems to be the most used approach. For simple Map-Reduce jobs the straightforward solution is Streaming but this solution is limited to text only input data files. For more complex jobs the solution should be Rhipe or RHadoop.

**References**

1. Ahas, R., and Tiru, M., (2013) Using mobile positioning data for tourism statistics: Sampling and data management issues, NTTS - Conferences on New Techniques and Technologies for Statistics, Bruselles.
2. Beyer, M., (2011), "Gartner Says Solving 'Big Data' Challenge Involves More Than Just Managing Volumes of Data". Gartner, available at http://www.gartner.com/newsroom/id/1731916, accessed on 25th March 2014.
3. Cleveland, William S., Guha, S., (2010), Computing environment for the statistical analysis of large and complex data, Doctoral Dissertation, Purdue University West Lafayette.
4. Dean, J., and Ghemawat, S., (2004), "MapReduce: Simplified Data Processing on Large Clusters", available at http://static.googleusercontent.com/media/research.google.com/ro//archive/mapreduce-osdi04.pdf, accessed on 25th March 2014.
5. High-Level Group for the Modernisation of Statistical Production and Services (HLG), (2013), What does "big data" mean for official statistics?, UNECE, available at http://www1.unece.org/stat/platform/pages/viewpage.action?pageId=77170614, accessed on 25th March 2014.
6. Holmes, A. (2012), Hadoop in practice, Manning Publications, New Jersey.
7. Mayer-Schönberger, V., and Cukier, K., (2012), "Big Data: A Revolution That Transforms How we Work, Live, and Think", Houghton Mifflin Harcourt.
8. Prajapati, V., (2013), Big data analysis with R and Hadoop, Pakt Publishing.
9. R Core Team, (2013), An Introduction to R, available at http://www.r-project.org/, accessed on 25th March 2014.
10. Ryan, A., (2012), Under the Hood: Hadoop Distributed Filesystem reliability with Namenode and Avatarnode, available at http://www.facebook.com/notes/facebook-engineering/under-the-hood-hadoop-distributed-filesystem-reliability-with-namenode-and-avata/10150888759153920, last accessed on 25th March, 2014.
11. Todorov, V. and M. Templ, (2012), R in the statistical office: Part 2, Development, policy, statistics and research branch working paper 1/2012., United Nations Industrial Development, 2012.
12. Todorov, V., (2010), R in the statistical office: The UNIDO experience. Working Paper 03/2010 1, United Nations Industrial Development. Available at: http://www.unido.org/fileadmin/user_media/Services/Research_and_Statistics/statistics/WP/WP_2010_03.pdf, accessed on 25th March 2014.
13. White, T., (2012), Hadoop: The Definitive Guide, 3rd Edition, O'Reilly Media.
14. Yahoo! Developer Network, (2014), Hadoop at Yahoo!, available at http://developer.yahoo.com/hadoop/, last accessed on 25th March, 2014.